\input psfig
\magnification=1200
\baselineskip=18pt
\def\lsim{<\kern-2.5ex\lower0.85ex\hbox{$\sim$}\ }
\def\rsim{>\kern-2.5ex\lower0.85ex\hbox{$\sim$}\ }
\def\ov{\overrightarrow}
\overfullrule=0pt
\noindent{Published in {\it Potentiality, Entanglement and 
Passion-at-a-distance - 
Quantum Mechanical Studies for Abner 
Shimony}, volume 2, edited by R. S. Cohen, M. Horne and J.
Stachel, 
(Kluwer, Dordrecht, Holland 1997), p. 31-52.}
\vskip .5cm

\centerline{\bf CLASSICAL AND QUANTUM PHYSICAL GEOMETRY} 
\vskip .5cm
\centerline{JEEVA S. ANANDAN}

\centerline{Departments of Physics and Philosophy} 
\centerline{University of South Carolina}
\centerline{Columbia, SC 29208, USA.} 
\centerline{and}
\centerline{Sub-Faculty of Philosophy, University of Oxford} 
\centerline{10 Merton St., Oxford OX2 4JJ, UK.}
\vskip .5cm

\centerline{ABSTRACT}

{\it The task of creating a quantum theory of gravity is compared 
with Einstein's creation of a relativistic theory of gravity. The 
philosophical and physical foundations of this theory are 
briefly reviewed. The Ehlers-Pirani-Schild scheme of operationally 
determining the geometry of space-time, using freely falling 
classical particle trajectories, is done using operations in an 
infinitesimal neighborhood around each point. The study of the 
free fall of a quantum wave suggests a quantum principle of 
equivalence. The principle of general covariance is clarified. The 
sign 
change of a Fermion field when rotated by $2\pi$ radians is used 
to 
argue for a quantum mechanical modification of space-time, which 
leads naturally to supersymmetry. A 
novel effect in quantum gravity due to the author 
is used to extend Einstein's hole argument to quantum gravity. 
This suggests a quantum 
principle of general covariance, according to which the 
fundamental laws of physics should be covariant under `quantum 
diffeomorphisms'. This heuristic principle implies that space-time 
points have no invariant meaning in quantum gravity.}

\vskip .5cm
\noindent
gr-qc/9712015

\vfil\eject

\noindent{INTRODUCTION: PHYSICS AND PHILOSOPHY}

\vskip .5cm
Many physicists and some philosophers hold the view that physics 
is an empirical science and that philosophers therefore have no 
place in it, except perhaps as historians. Abner Shimony has over 
the years opposed this narrow view of both physics and 
philosophy. One day he aptly summarized his distinguished roles 
in physics and philosophy by describing himself to me as a 
`natural philosopher' and an `experimental metaphysician'. He 
thereby emphasized the role of physics as natural philosophy and 
the relevance of philosophical principles to an experimental 
science such as physics. I was always in full agreement with this 
view. It is particularly relevant today because of the problem of 
quantizing gravity which has eluded the conventional methods of 
physicists and seems to call for a major paradigm shift.
It seemed to me therefore that a fitting contribution to the 
Volume in Abner's honor would be to describe some work I have 
done towards applying philosophical principles to the task of 
quantizing gravity, which may be the most difficult and deepest of 
all the unsolved problems in theoretical physics today. 

If we compare this task with the creation of quantum theory 
during the early part of this century, which led to a major 
paradigm shift, we find that there was a great deal of 
experimental evidence which physicists such as Planck, Einstein, 
Bohr, De Broglie, Heisenberg and Schr\"odinger could make use of 
in 
order to create quantum theory. This theory is so rich and counter 
intuitive that it would not have been possible for us, mere 
mortals, to have dreamt it without the constant guidance provided 
by experiments. This is a constant reminder to us that nature is 
much richer than our imagination. But there is no direct 
experimental evidence today on quantum gravitational 
phenomena which could guide us similarly in the construction of a 
quantum theory of gravity. So, we are left with the need to apply 
clever mathematical techniques, as in the case of superstring 
theory, or to apply philosophical principles, as in the present 
article, or both in order to create quantum gravity from almost 
nothing.

On the other hand, if we compare this task with another major 
paradigm shift of this century that accompanied the creation of 
general relativity, we find that the latter occurred with almost no 
guidance from experiment. This, I believe, was largely because of 
the genius of Einstein in judiciously applying philosophical 
principles and geometrical concepts to Newtonian gravity and 
special relativity, which led to the discovery of the deeper theory 
of general relativity which contained the first two theories as 
approximations. 
In section 1, I shall briefly describe this and argue in favor of 
following in Einstein's footsteps again.
\vskip .5cm
\noindent{1. RELATIVIZING AND QUANTIZING 
GRAVITY}
\vskip .5cm

After the discovery of special relativity by Lorentz, Poincare, and 
Einstein, there was the problem of ``relativizing gravity'', 
analogous 
to the problem of ``quantizing gravity'' which exists today. It was 
clear that Newtonian gravity was incompatible with special 
relativity and it was necessary to replace it with a relativistic 
theory of 
gravity. 
While several attempts were made to 
do this, Einstein succeeded in constructing such a theory because 
he 
used i) the {\it 
geometrical} reformulation of {\it special relativity} by 
Minkowski, and 
ii) the {\it operational} 
approach of asking what may be learned 
by probing gravity using {\it classical} 
particles. 

An important ingredient in (i) was Einstein's realization that the 
times in the different inertial 
frames, $t$ and $t'$, in the Lorentz transformation were on the 
same footing. This made the Lorentz group of transformations a 
true symmetry of physics. Minkowki then constructed a space-
time geometry by means of the metric that is invariant under 
the Poincare group of tranformations that is generated by the 
Lorentz 
transformations and translations acting on space-time. So, the {\it 
interpretation} Einstein gave to special relativity, whose basic 
equations were already known 
to Lorentz and Poincare, was crucial to the subsequent work of 
Minkwoski. It enabled 
Einstein to get rid of the three dimensional ether, and 
thereby pave the way for the 
introduction of the four dimensional `ether', called space-time, by 
Minkowski. 

By means of 
(ii), Einstein concluded that the aspect of Newtonian gravity which 
should be retained when this theory is modified is the 
equivalence principle. This principle is compatible with special 
relativity locally. This may be seen from the physical formulation 
of the strong equivalence principle according to which in the 
Einstein elevator that is freely falling in a gravitational field the 
laws of special relativity are approximately valid. But this 
principle allowed for the 
modification of special relativity to incorporate gravity as 
curvature of space-time. 

Today we find that general relativity, the beautiful theory of 
gravity which Einstein discovered in this way, is incompatible 
with quantum theory. Can we then adopt a similar approach? This 
would mean that we 
should use 1) a {\it geometrical} reformulation of {\it quantum 
theory}, and 2) an {\it 
operational} approach of asking what may be learned 
by probing gravity using {\it quantum} 
particles. 

As for (1), the possibility of using group elements as `distances' in 
quantum theory, analogous to space-time distances in classical 
physics, was studied previously [1]. For a particular quantum 
system, the corresponding representations of these group 
elements may be used to relate points of the projective 
Hilbert space, i.e. the set of 
rays of the Hilbert space, which is the quantum generalization of 
the classical phase space [2]. Recent 
work on protective observation of the quantum state has shown 
that the points of the projective 
Hilbert space 
are real, in the sense that they could be observed by 
measurements on an individual system, 
instead of using an ensemble of identical systems [3].

As for (2), the question is whether the motion of a quantum 
system in a gravitational 
field enables us to identify the aspect of general relativity 
which must be preserved when this theory is replaced by a 
quantum theory of gravity, i.e. the quantum analog of the 
equivalence principle. 
I shall formulate such a principle, in this article. 

In section 2, I shall review the classical equivalence 
principle and its use by Ehlers, Pirani and Schild (EPS) to 
determine the geometry of space-time from the trajectories of 
freely falling particles. I shall then provide a new formulation of 
the equivalence principle, in section 3, in terms of the symmetry 
group acting in the first order infinitesimal neighborhood around 
each point. This modified equivalence principle is simpler and 
leads to the geometry more naturally than the EPS scheme. Also, it 
shows the connection between the different structures studied by 
EPS. Moreover,
the EPS scheme breaks down when we go to quantum theory 
because the particles do not have trajectories (except in the Bohm 
interpretation of quantum theory in which the trajectories 
assigned to the particles are for the most part unobservable and 
therefore cannot be used to obtain the geometry). But the 
modified equivalence principle has a smooth transition to 
quantum theory. 

This will be shown in section 4, where the 
objective will be to do the quantum mechanical version of the EPS 
scheme, i.e. to determine the geometry using wave motion instead 
of particle motion. A quantum weak equivalence principle and a 
quantum strong equivalence principle will be formulated. It may 
be noted that Einstein's equivalence principle, which he 
discovered in 1908, was largely a philosophical princple until the 
mathematical construction of general relativity. Similarly, the 
present quantum equivalence principles are largely philosophical, 
and would probably remain so until the construction of quantum 
gravity.

The principle of general covariance used by Einstein in his 
discovery of general relativity is studied in detail in section 5. 
The role of coordinate systems and symmetries is clarified. 
Einstein's hole argument is examined and the distinction between 
passive and active transformations is abolished.

In section 6, a novel effect due to the quantum superposition of 
two geometries on the wave function of a test particle is 
described. This 
effect is invariant under a quantum diffeomorphism that 
transforms different geometries differently. This freedom 
suggests that the points 
of space-time have no 
invariant meaning. So, there seems to be a 
need to get rid of the four dimensional `ether', 
namely space-time, in order to 
incorporate the quantum diffeomorphism symmetry into 
quantum gravity. The covariance of the laws of physics under 
these quantum diffeomorphisms is formulated as a new principle 
of quantum general covariance. 

\vskip .5cm
\noindent{2. THE CLASSICAL EQUIVALENCE 
PRINCIPLE} 

\noindent{AND THE 
EHLERS-PIRANI-SCHILD SCHEME}
\vskip .5cm

The classical weak equivalence principle (WEP), due 
to Galileo and 
Einstein has two aspects to it: In a space-time manifold with a 
pure 
gravitational field, a) the possible motions of all freely falling test 
particles are the same, and b) at any point $p$ in space-time, 
there 
exists a neighborhood $U(p)$ of $p$ and a coordinate system 
$\{x^\mu , \mu=0,1,2,3\}$, such that the trajectories of {\it every} 
freely falling test particle through $p$ satisfies [4]
$${d^2x^\mu\over d\lambda^2}=0, \eqno(2.1)$$
{\it at p} for a suitable parameter $\lambda$ along the trajectory. 
This is the 
local form of the law of inertia and 
the above coordinate system is said to be locally inertial at $p$. 
The 
condition (b) is a special property of the gravitational field, not 
shared by any other field. For example, in an electromagnetic field 
test particles with the same charge to mass ratio would satisfy 
(a) but not (b). (The Lorentz 4-force is 
proportional to the electromagnetic field strength which, being a 
tensor, cannot be coordinate transformed away unlike the 
connection 
coefficients.)

Using (b), for massive and massless particles, it was shown by 
Ehlers, Pirani and Schild (EPS) [4], based on the earlier work of 
Weyl, that there exists an affine connection 
$\omega$ such that the trajectories of freely falling test particles 
are 
affinely parametrized geodesics with respect to it. I shall now 
present their arguments more clearly by means of operations in 
an infinitesimal neighborhood around each point, instead of using 
differential equations. The use of this neighborhood, which will 
be defined shortly, also will pave the way for an improved 
version of the classical equivalence principle by means of the 
symmetry group in this neighborhood, in section 3. The latter 
principle will be seen 
to have a smooth transition into quantum physics, 
unlike the equivalence principle as formulated by Galileo, Einstein 
or EPS.

Suppose 
$\epsilon 
= {d\over L}$, where $d\sim$ linear dimensions of $U(p)$ and 
$L\sim$ radius of curvature obtained from the curvature 
components of this connection, all lengths being measured in the 
above coordinate system, and we can neglect second orders in 
$\epsilon$. Such a neighborhood will be called a first order 
infinitesimal neighborhood of $p$, and denoted by $U_\epsilon 
(p)$. 
Using the geodesic deviation equation, it may be shown that the 
velocities of the freely falling test particles in $U_\epsilon(p)$ are 
constant in an appropriately chosen coordinate system. This is a 
stronger form of the WEP than its usual statement given above, 
and 
will be called the {\it modified classical weak equivalence} 
principle. It is 
valid in Newtonian gravity as well as Einsteinian gravity. 

Let us now look at the different geometrical structures that arise 
in $U_\epsilon(p)$ directly from the motions of particles, instead 
of assuming an {\it a priori} metric as in the above analysis. 

Specifying the unparametrized geodesics in $U_\epsilon (p)$ gives 
it a projective geometry. Now the trajectories of massive freely 
falling particles are time-like geodesics. But the collection of such 
trajectories that pass through a given point contain as their 
boundary, the collection of null geodesics at that point. The 
tangent vectors to these null geodesics constitute the null cone at 
that point. Specification of the null cone at each point in space-
time is the same as specifying the conformal structure of space-
time. 

\centerline{\psfig{figure=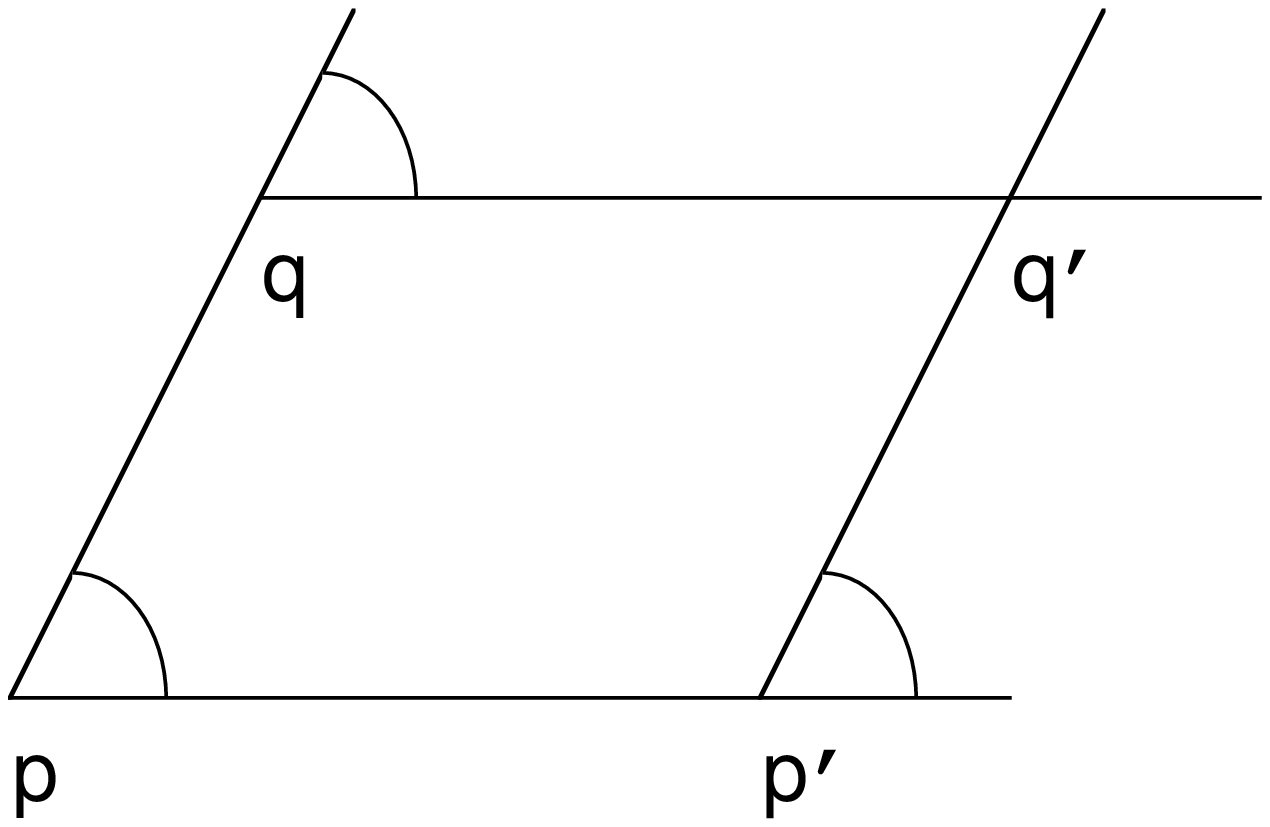,height=3.0in,width=4.88in}}

\noindent
Figure 1. Determination of the affine structure using the projective 
and conformal stuctures. The projective structure determines the 
lines and the conformal sturcture determines the equality of the 
angles indicated in the infinitesimal parallelogram $pp'q'q$. Then 
$p'q'$ may be regarded as the prallel transport of $pq$, which gives 
rise to the affine stucture.

\vskip .5cm

The projective structure determines the ``straight lines'', or simply 
``lines'' (the preferred curves of the projective structure) and the 
conformal structure determines angles in $U_\epsilon (p)$. Using 
these two concepts, an infinitesimal parallelogram may be 
constructed in 
$U_\epsilon (p)$ as follows: Let $p'$ and $q$ be two points that 
are 
in $U_\epsilon (p)$ and distinct from $p$. There exist two unique 
preferred curves of the projective structure ( ``lines'' in 
$U_\epsilon 
(p)$) passing through $p,p'$ and $p,q$ (Fig. 1). Let $V$ be the two 
dimensional vector space spanned by the tangent vectors to these 
two curves at $p$. The ``lines'' through $p$ that have the vectors 
in 
$V$ as tangent vectors form a two dimensional surface in 
$U_\epsilon (p)$ which will be called a ``plane''. Choose the unique 
point $q'$ on this 
plane such that the line segment $p'q'$ makes the same angle as 
$pq$ with $pp'$, and $qq'$ makes the same angle as $pp'$ with 
$pq$, 
as determined by the conformal structure. Then $qq'$ will be said 
to 
be parallel to $pp'$, and $p'q'$ will be said to be parallel to $pq$. 
Therefore, $pp'q'q$ is an infinitesimal parallelogram. It is 
emphasized that this construction does not use a metric.
We now have an affine geometry in
$U_\epsilon (p)$, because an affine geometry is a projective 
geometry together with the concept of parallelism. It is possible 
now to parallel transport a vector along an arbitrary curve as 
follows: Given two infinitesimally separated points $p$ and $p'$ 
on the curve such that $p'$ is inside $U_\epsilon (p)$, construct an 
infinitesimal parallelogram $pp'q'q$, with the direction of $pq$ 
being arbitrary, using the 
projective and conformal structures, as described in the previous 
paragraph. Then $p'q'$ is the parallel 
transport of $pq$. Then by 
suitable rescaling, any tangent vector at $p$ in the direction of 
$pq$ may be parallel transported to $q$ to be in the direction of 
$qq'$. Note that this prescription fixes both the direction and the 
length of the transported vector as it should be under parallel 
transport, but without requiring a metric do so. Now that parallel 
transport for an infinitesimal displacement is known, it is possible 
to parallel transport an arbitrary vector along the entire curve, 
which is arbitrary except that it is piecewise differentiable.
This defines an affine connection.

It may be noted that this affine connection is torsion free. This is 
because in the presence of torsion infinitesimal parallelograms do 
not 
close in general. Whereas the above affine connection is defined so 
that infinitesimal parallelograms always do close. 

The affine connection together with the conformal structure is 
called a Weyl structure. In a Weyl space-time, it is possible to 
compare the lengths of two measuring rods at a given space-time 
point using the conformal structure. Also, it is possible to parallel 
transport either of them, using the affine connecton, so that the 
rod remains the `same' during this process. (Cf. the opposite sides 
$pq$ and $p'q'$ of the above infinitesimal parallelogram, which 
are equivalent with respect to the affine geometry and therefore 
the `same'.) But when this rod is parallel transported around a 
closed curve it would in general undergo a rotation (Lorentz 
transformation) and an elongation or contraction compared to the 
rod that was left at the original point. 

The Lorentz transformation of a 4-vector under parallel transport 
around a closed curve is due to the space-time curvature which 
represents the gravitational field. Weyl tried to identify the 
change of length of the vector as being due to the electromagnetic 
field. This shows a lack of operationalism in Weyl's approach to 
the electromagnetic field, unlike Einstein's approach to the 
gravitational field via the equivalence principle obtained by 
probing the gravitational field with a classical particle. If 
we characterize the electromagnetic field by what it does to a 
charged 
probe that is used to measure the field, we find that the field does 
not cause any change in length. The field exerts forces on a 
classical charged particle, and it produces a phase factor on the 
wave function of a quantum particle. I shall deal with the latter 
aspect in more detail in section 4.

There is no experimental evidence at all for the above mentioned 
change in length postulated by Weyl with or without the 
electromagnetic field. It is therefore reasonable to suppose that 
space-time is Riemannian, i.e. it is a special case of Weyl space-
time in which a vector parallel transported around a closed curve 
may come back rotated but without any change of length. EPS 
make this as an additional postulate in order to obtain the 
Riemannian structure of space-time.

\vskip .5cm
\noindent{3. THE MODIFIED STRONG EQUIVALENCE PRINCIPLE}
\vskip .5cm
There are two shortcomings in the 
EPS scheme, described in section 2. First, by using freely falling 
particle trajectories that satisfy the equivalence principle, EPS 
obtain naturally the Weyl structure and not the Riemannian 
structure. The additional postulate they make to obtain 
Riemannian geometry is ad hoc and is not 
suggested naturally by the operational procedure they adopted. 
Secondly, they consider several geometrical stuctures, and the 
connection between 
them appear mysterious. This seems to call for a simpler and 
more unifying principle.

I shall now give a new formulation of the equivalence 
principle, 
which does not have these shortcomings.
The formulations of WEPs, given in section 2, may be stated 
using 
only 
an affine connection and do not require a metric. In 
$U_\epsilon$, the 
affine structure defined by this connection has as its symmetry 
group the 
affine group $A(4)$ that is generated by the general linear 
transformations and translations in a $4$ dimensional real vector 
space. In the non relativistic limit, as the null cones `flatten',  
$A(4)$ remains the symmetry group. 

When the translational subgroup of $A(4)$ acts on a given point, 
the orbits are geodesics, which are the trajectories of freely falling 
particles. These curves satisfy the condition (2.1), and fulfill the 
usual formulation of the equivalence principle. Physically 
speaking, the translational invariance which exists in the absence 
of external forces implies, via Noether's theorem, the conservation 
of energy-momentum. Satisfying (2.1) is due to the local 
constancy of the energy-momentum of the freely falling particle. 
These preferred curves define the projective structure. So, we see 
here the 3-fold connection between the symmetry group, the 
geometry and the physics, in this particular case of the 
equivalence principle.

In classical physics, 
the {\it interactions} between the particles 
restrict the symmetry group in $U_\epsilon$ to the 
inhomogeneous 
Galilei group (non relativistic physics), or the Poincare group $P$ 
(relativistic physics), which are both subgroups of $A(4)$. The 
existence of this residual symmetry group in $U_\epsilon$ is a 
form of the classical strong equivalence principle (SEP) valid for 
relativistic and non relativistic gravity. I shall call this the {\it 
modified classical SEP}.

In this way, non flat 
space-time geometry may also in 
some sense be brought into the frame-work of Felix Klein's 
Erlanger program according to which a 
geometry 
is determined as the set of properties invariant under a symmetry 
group [1]. 

As mentioned above, the translational subgroup acting on 
$U_\epsilon (p)$ of the Poincare group determines the projective 
structure. The Lorentz subgroup leaves invariant the null cone at 
each point $p$ and therefore determines the conformal structure. 
So, the relationship between these two structures can now be 
understood algebraically in terms of the relationship between 
these two subgroups of the Poincare group. 

The modified SEP also can be extended to wave motion. The 
particle trajectories which EPS used are obtained in the geometric 
optics limit of the quantum wave. In this limit, the information 
contained in the phase of the wave function is lost. Once this 
phase information is restored, the compatibility between the 
metric and the connection 
which EPS introduced, in order to specialize the Weyl structure 
resulting 
from the projective and conformal structures to the Riemannian 
structure, naturally follows. 

The metric compatibility follows from the fact that in 
quantum theory there is a natural frequency $\omega$ associated 
with a mass $m$ given by
$$mc^2 =\hbar \omega \eqno(3.2)$$
which acts as a clock. Using the distances along time-like curves 
measured by this clock and light signals, whose motion is 
determined by the conformal structure, it is possible to determine 
the metric, as shown long ago by Synge. And 
$m^2=\eta^{ab}P_aP_b$ is a Casimir operator of the Poincare group 
which means that it commutes with every element of this group. 
It will be shown in the next section that this implies that space-
time is Riemannian.

\vskip .5cm
\noindent{4. THE QUANTUM EQUIVALENCE PRINCIPLE}
\vskip .5cm

What fundamental aspects about the gravitational field may be 
learned if it is probed with quantum particles, instead of with 
classical particles as in the above treatment? It was shown that 
the 
evolution of a freely falling wave function is given, in the WKB 
approximation, by the action on the initial wave function by the 
operator [5]
$$\Phi_\gamma =  Pexp[- i\int_{\gamma} \Gamma_\mu dx^\mu 
],\eqno(4.1)$$
where 
$$\Gamma_\mu ={\theta_\mu}^a 
P_a +{1\over2}{{{\omega}_\mu}^a}_b {M^b}_a . \eqno(4.2)$$
which will be called the gravitational phase operator. Here the 
energy-momentum operators $P_a$ and the angular 
momentum operators ${M^b}_a, a,b = 0,1,2,3$ generate the 
covering 
group of the Poincare group $\tilde P$ that is a semi-direct 
product 
of $SL(2,C)$ and space-time translations $R(4)$. The fact that 
mass 
$m$ is a good quantum number in curved space-time and $m^2$ 
is 
a 
Casimir operator of $P$ already suggests that $P$ is relevant in 
the 
presence of gravity. 

For every space-time point $p$, let $H_\epsilon (p)$ be the Hilbert 
space of wave functions in $U_\epsilon (p)$ in which $\tilde P$ 
acts. 
Owing to the linearity of the action of (4.1), it determines also the 
evolution of any freely falling wave packet which can be 
expanded 
as a linear combination of WKB wave functions, provided the size 
of 
the wave packet is small compared to the radius of curvature, i. e. 
it 
is contained primarily inside $U_\epsilon$ at each point along 
$\gamma$ which may be chosen to be along the center of the 
wave 
packet. This will be called the {\it quantum weak equivalence 
principle}, 
because (4.1) is a Poincare group element 
independent of the freely falling 
wave packet. In this respect, it is like the classical WEP according 
to which the affine connection determined is independent of the 
test 
particle used.

In quantum physics, because the wave packet must necessarily 
have 
some spread, the WEP cannot be formulated by particle 
trajectories 
as in conditions (a) and (b) in section 3, and it is necessary to 
use at 
least 
the neighborhood $U_\epsilon$. Indeed (4.1) was obtained [5] 
using 
the Klein-Gordon [6] and Dirac equations [7] which are covariant 
under 
$\tilde P$ in $U_\epsilon$. So, in quantum physics there is a close 
connection between the WEP, as formulated above, and SEP 
according 
to which $\tilde P$ is the symmetry group of all laws of physics in 
$U_\epsilon$. It is well known that (a) cannot be valid in 
quantum physics, because the motions of wave functions depend 
on their masses [8]. 
But the modified classical WEP and the classical SEP formulated in 
sections 2 and 3
have the advantage that they have a smooth transition to 
quantum 
physics. 

The above approximate concepts may be made mathematically 
precise as follows: Each neighborhood $U_\epsilon (p)$ may be 
identified with the tangent space at $p$ regarded as an affine 
space. 
The motions of freely falling test particles relate affine spaces 
associated with two neighboring points by a linear transformation 
and a translation, generated by $P_a$. 
This gives a natural connection on the affine bundle [9] over 
spacetime which is a principal fiber bundle with $A(4)$ as the 
structure group. This is the connection used above to express the 
modified
classical WEP. The quantum WEP requires the Poincare subbundle 
with $\tilde P$ (to admit Fermions) as the structure group. Then 
(4.2) defines a connection in 
this 
principal fiber bundle. The gravitational phase operator (4.1) 
parallel transports with respect to this connection along the curve 
$\gamma$. The above Hilbert space bundle, that is the union of 
$H_\epsilon (p)$ for all space-time points $p$, is a vector 
bundle associated to this principal fiber bundle with a connection 
that is the representation (4.2) in this Hilbert space.  

The curvature of the above connection is the Poincare Lie algebra 
valued 2-form
$$F =d\Gamma+\Gamma\wedge\Gamma =
Q^aP_a +{1\over2}{R^a}_b {M^b}_a ,\eqno(4.3)$$
where, on using) and the Lie algebra of the Poincare group, 
$$Q^a=d\theta^a + {\omega^a}_b\wedge\theta^b ,{R^a}_b 
=d{\omega^a}_b+{\omega^a}_c\wedge{\omega^c}_b .\eqno(4.4)$$
which are called respectively the 
torsion and the linear curvature.
If the wave equation used to obtain (4.1) did not contain torsion, 
then the torsion in (4.3), of course, is also zero. However, the 
above modified classical WEP and the quantum WEP make it 
natural 
to have torsion and suggest that if the torsion is zero then there 
should be a 
good physical reason for it.

Suppose $\gamma$ is a closed curve. Then (4.1) is a holonomy 
transformation determined by the above affine connection. 
The importance of (4.1) may also seen by comparing it with 
the corresponding phase factor for electromagnetism:
$$exp\left ( -i\int eA_\mu dx^\mu\right)  \eqno(4.5)$$
which is an element of the U(1) gauge group, where $A_\mu$ is 
the 4-
vector potential. It was pointed out by Yang [10] that the 
importance of (2), which appears in the wave function of a 
particle with charge $e$, was recognized by Schrödinger [11] in 
1922, 
in his study of Weyl's gauge theory, four years before he 
introduced the wave function. The question then arises whether 
(4.1) is similarly the "shadow" of some important yet to be 
discovered concept in quantum gravity. In any event, the analogy 
between (4.1) and (4.5) implies that gravity may be regarded as a 
gauge field in the spirit of Chen Ning Yang's integral formulation 
of gauge field [12].

Evaluating (4.1) for a closed curve spanning an infinitesimal 
area $d\sigma ^{\mu\nu}$ gives
$$\Phi_\gamma =1 + {i\over 2} (Q_{\mu\nu}{}^a Pa + {1\over 2} 
R_{\mu\nu}^{ab} 
M_{ba})d\sigma ^{\mu\nu} , \eqno(4.6)$$
where $Q^a$ and ${R^a}_b$ are respectively the torsion and linear 
curvature. While this makes it natural to introduce torsion into 
gravity, there have been no experimental tests so far to test the 
presence of torsion, or to put an upper limit on it. 

But motivated 
by this result, I obtained as an exact solution the most general 
stationary cosmic string containing torsion [13], by solving the 
simplest generalization of Einstein's gravitational field equations 
to include torsion. This is the gravitational analog of the solenoid 
in electromagnetism which produces the Aharonov-Bohm (AB) 
effect [14]. The gravitational AB effect due to the phase factor (1) 
is 
considerably richer [15]. Also, the solution I obtained is of interest 
in astrophysics because of the possible role of cosmic strings in 
galaxy formation, which is an important problem in explaining the 
observed universe. 

It follows from (4.6) that in the absence of gravity in a simply 
connected region (4.1) is path independent. I shall take the 
equivalent 
statement that  the path dependence of (4.1) implies gravity as 
the 
{\it definition} of the gravitational field 
even when the region is not 
simply connected. This definition makes the converse of this 
statement also valid. So, by probing 
gravity using quantum mechanical systems, without 
paying any attention to gauge fields, gravity may be 
obtained naturally as a 
Poincare gauge field.

Comparing now the present scheme with the EPS scheme, which 
uses particle motion to obtain the geometry of space-time, the 
present scheme, which uses wave motion instead, does not need to 
bring in 
anything external in order to obtain compatibility of metric and 
connection. To see this consider two beams which go along two 
different paths from one space-time point $A$ to another point 
$B$. The metric along each beam is determined by the Casimir 
operator. But (4.1), which determines the evolution of each beam, 
being an element of the Poincare group, commutes with the 
Casimir operator $m^2$. Therefore this Casimir operator remains 
the same as it is transported along each beam and hence, using 
the phase as a clock, the two identical clocks along the two beams 
are in agreement after the two beams meet.

An advantage of this point of view is that it also provides a 
unified description of gravity and gauge fields. If a wave function 
is interacting not only with the gravitational field but also other 
gauge fields, then its propagation in the WKB approximation is 
given by the action of an operator of the form (4.1) with
$$\Gamma_\mu ={\theta_\mu}^a 
P_a +{1\over2}{{{\omega}_\mu}^a}_b {M^b}_a +{A_\mu}^jT_j , 
\eqno(4.7)$$
where ${A_\mu}^j$ is the Yang-Mills vector potential and $T_j$ 
generate the gauge group $G$. So, (4.1) now is an element of the 
entire symmetry group, namely $\tilde P\times G$. Thus, unlike 
the 
classical WEP, the quantum WEP naturally extends to incorporate 
all gauge fields. 

The above fact that the observation of all the fundamental 
interactions 
in nature is via elements of the symmetry group suggest a 
symmetry ontology. By this I mean that the elements of 
symmetry group are observable and therefore real. Moreover, the 
observables such as energy, momentum, angular momentum and 
charge, which are usually observed in quantum theory are some 
of the generators of the above symmetry group. Observation 
always requires interaction between the observed system and the 
apparatus. Ultimately, these interactions are mediated by gravity 
and gauge fields, which act on the matter fields through elements 
of 
the symmetry group. I therefore postulate that the only 
observables 
which can actually be observed are formed from the generators of 
symmetry group, which according to our current understanding of 
physics are generators of $\tilde P\times G$.

\vskip .5cm
\noindent{5. CLASSICAL GENERAL COVARIANCE AND 
SPACE-TIME 
POINTS}
\vskip .5cm

It was mentioned in section 1 that historical lessons from 
Einstein's relativization of gravity may be useful 
in the quantization of gravity. An important step in Einstein's 
journey towards general relativity, apart from the principle of 
equivalence already dealt with in sections 2.2-4, was his 
discovery of the {\it principle of general covariance}. Unlike the 
principle of equivalence whose importance, in suggesting the 
incorporation of gravity as curvature of space-time, was 
realized by Einstein as early as 1908, he did not feel comfortable 
with general 
covariance. Indeed he first rejected this principle in 1913 on the 
basis of his `hole argument' which will be discussed later in this 
section. This delayed the construction of general relativity by two 
years. His eventual resolution of the hole argument in favor of 
accepting general covariance enabled him to write down soon 
afterwards the gravitational field equations which overthrew 
Newtonian gravity, after its reign of two and a half centuries.

In view of the great confusion which surrounded and still 
surrounds the principle of 
general covariance and the role of coordinate systems among 
many physicists and philosophers, including Einstein, it would be 
worthwhile to examine it in some detail, as I shall do now. In the 
next section, I shall formulate a new principle of quantum general 
covariance which I hope would be similarly useful in constructing 
a quantum theory of gravity.

In special relativity, it was believed that there was a real, 
objective space-time manifold, the set of space-time points with a 
four dimensional 
Euclidean topology and differentiable structure. This manifold 
is simply connected and is endowed with an a priori, fixed 
Minkowski metric. I shall 
call this the absolute Minkowski metric, to distinguish it from 
other Minkowski metrics on this manifold which will be discussed 
shortly. Its curvature 
$$ R_{\mu\nu}{}^\rho{}_{\sigma} =0 \eqno(5.1)$$
everywhere. Conversely, the metric of Lorentzian signature that 
satisfies (5.1) everywhere in a simply connected space-time 
must 
necessarily be a Minkowski metric. However, the latter metric is 
not unique. This is 
readily seen from the fact that (5.1) is generally covariant, by 
which is meant that any diffeomorphism on space-time leaves the 
form 
of (5.1) unchanged. Therefore, given any metric that is a solution 
of (5.1), any diffeomorphism on space-time maps it into another 
metric 
which is also a solution of (5.1). Or, to put it even more trivially, 
the new metric is isometric, by definition, to the old metric and 
therefore describes the same flat geometry. It follows that this 
space-time has an 
infinite number of Minkowski metrics which are all solutions of 
(5.1). 

Hence, according to the present ontology of space-time, {\it giving 
the absolute Minkowski
metric on space-time has more information than giving (5.1) 
because it 
singles out one of the infinitely many possible Minkowski metrics, 
that are solutions of (5.1), as the actual metric}. Later, I shall use 
Einstein's hole argument to change this ontology, which will lead 
to the rejection of the above statement in italics. Then the latter 
statement would become analogous to Newton's attempt to 
introduce an absolute space even though Newton's laws are 
covariant under Galilei boosts.

The space-time manifold, together with the absolute Minkowski 
metric on it will be denoted by $M$. On $M$, which is our `arena', 
there are also matter fields, classical or quantum. These are 
`painted' on $M$ by which I mean that they are appropriately 
differentiable functions of $M$ that do not distort the 
(Minkowski) geometry of $M$. (I.e. treating special relativity as a 
limiting case of general relativity, the back reaction of the matter 
fields on the space-time geometry is neglected). I shall denote 
$M$ together with the matter fields on it that satisfy the laws of 
physics, and which are just as real as the points of $M$, by $M^*$. 
So, $M^*$ is a mathematical representation of a possible universe.

To focus our ideas, consider the classical electromagnetic field 
$F_{\mu\nu}$ which satisfies the Maxwell's equations
$$F^{\mu\nu}{}_{;\nu} = j^\mu , F_{[\mu\nu ,\rho] }=0, 
\eqno(5.2)$$
where $j^\mu$ is the current density. Here (5.2) is written in an 
{\it arbitrary coordinate system}, with ${;\nu}$ representing the 
covariant derivative using the Christoffel connection formed from 
the metric coefficients in this coordinate system. Such general 
coordinates are sometimes called curvilinear coordinates to 
distinguish them from the Minkowski coordinates in which the 
metric coefficients take the usual Minkowski form 
$\eta_{\mu\nu}$. They 
are useful for solving particular problems. E.g. if there is spherical
symmetry then it is {\it convenient} to use spherical polar 
coordinates.
 
Two types of 
transformation in $M$ may be distinguished. One is a {\it passive} 
transformation, which is a coordinate transformation amounting to 
a mere relabeling of the points of $M$. The other is an {\it active} 
transformation that is a diffeomorphism of $M$ onto itself, while 
the coordinate system is kept fixed. Both transformations leave 
(5.1) covariant. But singling out the absolute metric makes them 
very different. The active transformations which leave the 
absolute metric 
on $M$
invariant is the Poincare group generated by the Lorentz 
transformations and the space-time translations. But passive 
transformations consist of the much larger group of 
diffeomorphisms. This is because the specification of a coordinate 
system requires only the differentiable structure, and therefore a 
change of coordinates need to keep only the differentiable 
structure invariant.

The transformations on $M$ also transform appropriately the 
matter fields on $M^*$, which are tensor or spinor fields on $M$. 
To specify spinor fields on $M$, it is necessary also to define a 
`vierbein' field which is a differentiable choice of local Lorentz 
frames. 
Operationally, the value of a spinor field at a space-time point is 
what would be observed by an observer using the local Lorentz 
frame that is the value of the vierbein field at the same point. 
Therefore, transformation of the vierbein field must be specified, 
in addition to the coordinate transformation, in order to determine 
the transformation of the spinor field. But for each Minkowski 
coordinate system, it is convenient to choose the corresponding 
vierbein field to be the coordinate basis. Then the transformations 
between the Minkoski coordinate systems, consisting of the 
Poincare group of transformations, automatically determine the 
transformations of the vierbein field, and hence of the spinor 
fields. 

A physical process in $M^*$ is defined to be a collection of matter 
fields which satisfy all the laws of physics, given by equations 
such as (5.2). Then the principle of special relativity due to 
Einstein may be stated as follows: Given any physical process in 
$M^*$, its transform by an active Poincare transformation of $M$ 
is also a physical process. I emphasize that this formulation 
assumes an absolute space-time $M$ {\it relative} to which these 
transformations produce new configurations. Later on I shall give 
up this assumption, which would necessitate defining a symmetry 
transformation as keeping something fixed.  

An active transformation of $M^*$ that transforms any physical 
process to another physical process will be called a {\it symmetry} 
of 
the laws of physics. It is easily shown that the set of symmetries 
form a group. Instead of first specifying a metric a priori in $M$ 
and requiring that the active transformations which leave it 
invariant are also symmetries of the laws of physics, we could 
start with the group $G$ of symmetries on $M^*$ and obtain the 
geometry as the set of properties invariant under $G$, in 
accordance with Klein's Erlanger program [1]. In fact, if we insist 
on the determination of the geometry operationally by means of 
physical processes, involving clocks, measuring rods, etc., the 
symmetries of the laws of physics must necessarily be the 
symmetries of the geometry. This led to the formulation of the 
principle of physical geometry in ref. [1] according to which the 
symmetry group $P$ of the laws of physics is strictly the same as 
the 
symmetry group $G$ 
of the geometry:
$$P=G . \eqno(5.3)$$
From this point of view, we cannot make 
the above philosophical distinction between $M$ that contains 
fixed absolute geometrical structures and $M^*$ that contains in 
addition variable, dynamical structures.

In general relativity, which superseded special relativity, (5.1) is 
replaced by the field equations
$$G^{\mu\nu}=8\pi T^{\mu\nu}, \eqno(5.4)$$
where $G^{\mu\nu}$ is the Einstein tensor formed from the 
curvature tensor in (5.1) and $T^{\mu\nu}$ is the energy-
momentum tensor. Then (5.4) imply via the Bianchi identities 
that
$$T^{\mu\nu}{}_{;\nu} = 0, \eqno(5.5)$$
which represents local conservation of energy-momentum. In 
addition, a prescription should be given for determining 
$T^{\mu\nu}$ as a function of the matter fields in a generally 
covariant manner. Then (5.5) incorporates the equations of 
motion 
for the matter fields [16]. This is because the local conservation of 
energy-momentum in the interactions between matter fields 
largely
determine their evolution. For example, if $T^{\mu\nu}$ is the 
energy-momentum tensor of the electromagnetic field then (5.5) 
incorporates (5.2). 

So, a remarkable feature of the Einstein-Hilbert field equations 
(5.4) is that, together with the prescription for $T^{\mu\nu}$ as a 
functional of the matter fields, it incorporates {\it all the laws of 
classical 
physics}, because it implies (5.5). On the other hand, without this 
prescription, (5.4) is a tautology because it is then merely a 
definition of $T^{\mu\nu}$ which automatically satisfies (5.5).

Another remarkable feature of (5.4) which makes general 
relativity fundamentally different from any previous classical 
theory is that it makes the metric dynamical. By a field being 
dynamical here I mean that the field is not given 
a priori as a fixed or absolute object but is
determined by solving the field equations. In special relativity, 
the objects of $M$ were fixed, while the additional objects of 
$M^*$ were dynamical. Similarly, we may make a preliminary 
division between these two types of objects in general relativity: 
Define by $M_G$, the space-time manifold with only its topological 
and differentiable structures. Let $M_G^*$ denote $M_G$ together 
with the metric and matter fields.

Now, (5.4) is generally covariant. So, the symmetries of (5.4) 
are the group of diffeomorphisms on $M_G^*$. This is 
in accordance with the above mentioned principle of physical 
geometry
because the differentiable structure is the
geometry that is invariant under the group of diffeomorphisms. 

We may now distinguish between two types of general 
covariance. 
First, it is possible to cast any law of physics in a generally 
covariant
form, which is due to
coordinates being {\it labels}. This reflects the unavoidable 
freedom to change the coordinate system by any passive 
diffeomorphism, without changing any of the structures whether 
they are absolute or dynamical. I shall call this passive general 
covariance. This exists in special and general relativity. Second, 
there is the just described
symmetry group of active diffeomorphisms in general relativity 
due to 
the space-time metric becoming dynamical. I shall call this 
active general covariance. This is analogous to the local gauge 
symmetry
in gauge theories which is related to the dynamical nature of 
gauge fields.
 
In the light of the above analysis, let us now examine Einstein's 
resolution of 
the hole argument [17]. In 1913, Einstein and 
Grossmann [18] considered the determination of the gravitational 
field inside a hole in some known matter distribution by solving 
the gravitational field equations. If these field equations have 
active 
general covariance, then there are an infinite number of solutions 
inside the hole, which are isometrically related by 
diffeomorphisms. I shall call these geometries Einstein 
copies. 

This is unlike the case of the determination of the electromagnetic 
field in special relativity as described above. The electromagnetic 
field is uniquely 
determined inside a hole of some known charge distribution in 
Minkowski space-time $M$ by solving Maxwell's equations . 
Uniqueness here means that the field is obtained as a unique 
function of the space-time points inside the hole. But as 
mentioned earlier, these points may always be relableled by 
doing 
a coordinate transformation which is reflected in the passive 
general covariance of (5.1).

The Einstein copies may however be regarded as different 
representations of 
the same objective physical geometry. This follows if a 
space-time point inside the hole is defined operationally as the 
intersection of the world-lines of two material particles, or 
geometrically by the distances along geodesics joining this point to 
material points on the boundary of the hole. Under a 
diffeomorphism, such a point in one Einstein 
copy is mapped to a 
unique point in another Einstein copy. Both points may then be 
regarded as different representations of the same physical 
space-time point or an event. 

So, we may identify as the same universe the equivalence class of 
all Einstein copies of $M_G^*$ that are related by active 
diffeomorphisms. This abolishes the distinction between passive 
and active general covariance. Because, after the above 
identification has been made, the physical points of space-time 
remain the same under passive and active diffeomorphisms, both 
of which represent {\it equally} a mere change of labels.

Alternatively, as some philosophers have done, it is possible to 
regard the real universe $M_G^*$ as embedded in an uncountably 
infinite set of mathematical copies of $M_G^*$. The active 
diffeomorphism freedom then enables us to move around this 
infinite set only one of which is real. This approach may also be 
taken in gauge fields by treating only one of the infinite 
equivalence class of gauge potentials related by gauge 
transformations to be the real gauge field. 

But I reject the latter approach for the following three reasons: 
First, as already mentioned in connection with the hole argument, 
the operational determination of space-time points by matter 
fields forces us to identify all these different copies of $M_G^*$ 
because the corresponding points of $M_G^*$ are determined by 
the same procedure. Second, the latter view requires that we 
distinguish between active and passive transformations, whereas 
in differential geometry there is no distinction. Also, eqs. 
(5.1-13) are mathematically covariant equally under passive and 
active transformations. So, the latter approach does violence to the 
close connection between geometry and physics which the present 
paper regards as desirable. Finally, the above identification is 
necessitated by the use of Occam's razor, first because it reduces 
the uncountably infinite Einstein copies to the original and second 
because it abolishes the distinction between passive and active 
general covariance as shown above.

To recapitulate, there are two important philosophical points in 
the above analysis of general relativity. First, $M_G$ is used as 
the arena for the dynamical fields including the metric. Second, 
the identification of the Einstein copies of $M_G^*$ into a single 
space, which I 
shall call $\bar {M_G^*}$, makes the passive and active 
transformations the same. Both these aspects may be carried over 
to special relativity, which may be regarded as the limiting case of 
general relativity corresponding to weak gravitational sources and 
the special case of a simply connected space-time topology. Then 
$M_G$, instead of $M$, is the arena and the metric on $M_G$ may 
be obtained by solving (5.1), which is just as generally covariant 
as (5.4). But the different copies of $M_G$ together with the 
metrics on them that are solutions of (5.1), but which are 
diffeomorphically related, are now identified to be the same 
Minkowski space-time. This replaces $M$, and will be denoted by 
$\bar M$.

The question now arises as to what the symmetries are in special 
relativity. Geometrically speaking, the answer depends on which 
structures are kept fixed. If the Minkowski metric is regarded as 
an absolute structure, then the symmetries are the 
transformations which leave the geometry of $\bar M$ invariant, 
namely the Poincare group of transformations. If only the 
differentiable structure is kept fixed then the symmetries are the 
ones which leave the geometry of $\bar M_G$ invariant, namely 
the group of diffeomorphisms. A symmetry 
transformation is now redefined to assume implicitly that the 
transformation is being 
performed relative to something which is kept fixed, which 
we may take to be a frame [19]. This may be made precise by 
defining 
the transformation physically on $\bar {M_G}^*$ by specifying the 
frame by matter fields as some sort of a grid. Then the 
transformation gives a different 
$\bar {M_G}^{*'}$ which cannot be identified with $\bar {M_G}^*$ 
as 
was done above with the Einstein copies. The transformation 
being a symmetry means that $\bar {M_G}^{*'}$ is allowed by the 
laws of physics.

Instead of keeping the frame fixed and transforming the physical 
process, we may perform a passive symmetry transformation by 
keeping the 
physical process fixed and transforming the matter fields 
constituting the frame (grid). But the new $\bar {M_G}^{*''}$ 
obtained this way is an Einstein copy of $\bar {M_G}^{*'}$. So, the 
two should be identified according to our principle. This abolishes 
the distinction between active and passive symmetry 
transformations.

An example of the symmetry depending  on which structure is 
kept fixed are the Maxwell's equations (5.2) in special relativity. 
They are generally covariant with respect to the 
differentiable structure, but are Poincare covariant in relation to 
the Minkowski 
metric. 

As another example, consider the formulation of Newtonian 
mechanics as
generally covariant Lagrange's or Hamilton's equations. This 
general
covariance is a source of confusion among physicists and 
philosophers and
needs to be clarified. With respect to the symplectic structure in 
phase
space, the latter equations are generally covariant. Because all 
coordinate
transformations (diffeomorphisms) of space leave the 
symplectic structure in phase space
invariant. But with resepect to the Euclidean metric of Newtonian 
physics
(an absolute as opposed to a dynamical structure) this general 
covariance
is purely formal because the diffeomorphisms, in general, do not 
leave the Euclidean metric 
invariant.
It is the Euclidean group on space, or the ten parameter Galilean 
group on space-
time, which
leaves the entire geometry invariant and therefore is the
symmetry in this case.

\vskip .5cm
\noindent{6. FERMIONIC NATURE OF SPACE-TIME}
\vskip .5cm

In sections 2 and 5, classical matter fields were used to establish 
the 
ontology of the space-time manifold in classical physics. But in 
quantum theory, there are also Fermionic fields which have no 
classical analog. The question arises as to what kind of geometry is 
obtained if we apply the same philososphical and physical 
principles 
used to construct the classical physical geometry, above, now to 
the 
Fermionic 
fields. I shall present in this section joint work I did with Yakir 
Aharonov which provides an answer to this question by enlarging 
space-time so that it has a fundamental Fermionic nature.

The discussion of the hole argument in section 5 strongly 
suggested 
that we must regard space-time points as having a relational 
meaning instead of an absolute meaning. (This is implicit in the 
above specification of the points inside the hole by their distances 
along geodesics from points on the boundary of the hole.) In this 
spirit, consider the relationship between two neighboring space-
time 
points $A,B$. This may be specified by the connecting vector 
$\ov{AB} =\epsilon v$, where $\epsilon$ is infinitesimal and $v$ 
is a 
physically observable tangent vector at $A$. Then if the 
connecting 
vector is rotated about $A$ by $2\pi$ radians in some 2-plane, 
$B$ 
will return to itself. 

Consider now two points $A,B$ joined by a connecting spinor 
$\sqrt{\epsilon} \psi$. If this spinor is rotated by $2\pi$ radians it 
returns with the sign changed. It was first thought that this sign 
change is not observable, in which case we should say that $B$ 
returns to itself as before. But Aharonov and Susskind [20] 
showed 
that this sign change is observable. So, we must conclude that 
$B$ returns to a different point $B'$. 

The necessity for regarding $B$ and $B'$ as distinct points may 
also 
be seen from the following two arguments. As mentioned, when a 
spinor field is rotated by $2\pi$ radians then it does not return to 
itself. If this field is defined on the usual space-time then rotating 
the space-time, instead, by $2\pi$ radians would bring it back to 
itself. This would make active transformations different from 
passive 
transformations. But in section 5 it was argued that we should 
abolish the distinction between active and passive 
transformations. 
This could be accomplished now only if we allow for a passive 
transformation on space-time that would rotate $B$ to a distinct 
point $B'$.

Moreover, according to the principle of physical geometry (5.3), 
the 
symmetries of the physics are the same as the symmetries of the 
geometry. Since the rotations by $2\pi$ and $4\pi$ radians are 
distinguishable for Fermionic fields, the same should be true of 
the 
corresponding rotations acting on space-time. So, again, space-time 
should be enlarged so that the latter two rotations acting on the 
enlarged space-time are distinct.

This distinction may be made by specifying the variable 
point $B$ with respect to the fixed point $A$ by means of of the 
connecting spinor $\psi$ introduced above. Now, $\psi e^{i\theta}$ 
varies continuously 
from $\psi$ to $-\psi$ as $\theta$ varies from $0$ to $\pi$. Hence, 
this defines a continuous transformation from $B$ to $B'$, which 
are 
connected to $A$ by the spinor. Indeed as $\theta$ varies from 
$0$ 
to $2\pi$, it is clear that each space-time point now moves around 
a 
circle, whose points cannot be classically distinguished. We 
postulate 
that this $U(1)$ group of transformations is a 
symmetry of the theory for the following reason. In quantum 
theory, 
continuous symmetries are more natural than they are in classical 
physics. This is because, if a given configuration is transformed 
into 
another configuration by a discrete symmetry, then in quantum 
theory we can form a continuum of linear superpositions of the 
two 
configurations permitting a continuous group of transformations 
that 
connect the original discrete symmetry to the identity.

We can turn this argument around in view of the symmetry 
ontology 
proposed at the end of section 4. We may regard the symmetry 
group to be ontologically prior to the linear structure of the 
Hilbert 
space. It may be that the linear structure in quantum mechanics is 
required in order that continuous symmetries can act on the space 
of 
states.

Denote the generator of the new $U(1)$ symmetry, above, by $R$. 
Then, $\exp (i\theta R)$ acts on the geometry as well as the 
matter 
fields. Thus, space-time is now fundamentally quantum 
mechanical. 
Consider the action of this group on a pair of Bosonic and 
Fermionic 
fields. When $\theta =\pi$, the corresponding transformation 
should 
introduce a  relative minus sign between the two fields. If the 
initial 
state of this pair of fields is represented by the column vector 
$(1,1)^T$ then the action of $R$ on this pair of fields is 
represented 
by ${1\over 2}\sigma_z$, with $\sigma_x , \sigma_y ,\sigma_z$ 
being the usual Pauli spin matrices. 

But to distinguish between the initial state $(1,1)^T$ and the 
transformed state $i(1,-1)^T$, corresponding to $\theta =\pi$, it is 
necessary to have another 
observable, say $Q={1\over 2}\sigma_x$, which has these two 
states 
as its 
eigenvectors. Then, according to the symmetry ontology of section 
4, 
the observable $Q$ should generate a symmetry. Now, $Q$ 
generates 
supersymmetry transformation 
between the Bosonic and Fermionic fields. The commutator of $R$ 
with $Q$ gives another supersymmetric generator 
$Q'={1\over 2}\sigma_y$. So, $Q,Q'$ and $R$ generate an $SU(2)$ 
algebra. 

We illustrate the above ideas now by means of a simple example 
of 
a pair of Bosonic and Fermionic degrees of freedom, which are 
generated from the vacuum by the creation operators $a^\dagger$ 
and $b^\dagger$, 
respectively. These operators satisify the commutation 
and anticommutation relations
$$[a,a^\dagger ] =1, \{ b,b^\dagger \} =1, b^2 =0=(b^\dagger)^2 
.\eqno(6.1)$$
The Hamiltonian is
$$H= E(a^\dagger a+b^\dagger b) , \eqno(6.2)$$
where $E$ is a constant. Then, we may take
$$Q={1\over 2}N^{-{1\over 2}} \left( b^\dagger a+a^\dagger 
b\right),
Q'={i\over 2}N^{-{1\over 2}} \left( b^\dagger a-a^\dagger b\right),
R = -i[Q,Q'] = -2iQQ' , \eqno(6.3)$$
where $N=a^\dagger a+b^\dagger b$. Then,
$$ Q^2 = Q'^2=R^2 = {1\over 4} I, \{ Q,Q'\} =0, \eqno(6.4)$$
where $I$ is the identity operator.
Clearly, $Q$ and $Q'$ generate supersymmetric 
transformations between Bosonic and Fermionic states. And, it is 
easily 
seen that $Q,Q'$ and $R$ commute with $H$ and are therefore 
symmetries of this model. Also, $Q,Q'$ and $R$ generate an 
$SU(2)$ 
group.

To conclude this section, the application of our general principles 
to 
the sign change of the Fermion field when it is rotated by $2\pi$ 
radians has not only led naturally to supersymmetry, but also has 
given a new symmetry generated by $R$. The details of this work 
will be published elsewhere [21]. 

\vskip .5cm
\noindent{7. QUANTUM GENERAL COVARIANCE}
\vskip .5cm

It is a curious fact that most approaches to quantum gravity use 
the classical space-time manifold as the arena, as in classical 
general relativity, but with quantum fields (including the metric 
fields) instead of classical fields defined on it as operator value 
functions. But in fact quantum mechanics is formulated in Hilbert 
space and it is not possible to determine the points of space-time 
using quantum mechanical states because of the uncertainty 
principle. Operationalists such as E. Wigner have used this to 
argue that space-time manifold is not meaningful in quantum 
theory. Moreover, since each gravitational field is associated with 
a space-time geometry, quantizing it makes this geometry 
indefinite. So, even the points of space-time may be indefinite.
I shall now describe a new effect which I obtained in quantum 
gravity [22,13] which suggests that this may be the case. 

Suppose that the 
gravitational field of a string is quantized so that different 
geometries corresponding to different angular momenta of the 
string may be superposed. Each of these geometries is flat in any 
simply connected region outside the string. So, their separate 
effects on any given simply connected region $U$ would be like as 
if there is no graviational field. However if we put a test 
particle in the region $U$, and the string is observed to be in a 
superposition of different angular momentum states then the 
wave function of the test particle would be affected. Its intensity 
has a variation in position due to the superposed geometries even 
though each of them is flat. 

This effect is surprising and novel from a physical and 
philosophical point of view. What concerns us here are only the 
philosophical aspects which I shall discuss now. The above effect 
depends on the relationship between the metric coefficients of the 
two superposed geometries in $U$, in the particular gauge in 
which the gravitational field is quantized. But since each of these 
two geometries has zero curvature, it is possible to have quantized 
in a gauge in which both metric coefficients are the same and 
have the Minkowski values in $U$. So, how could their 
superposition affect the wave function of the particle? 

The pair of superposed geometries in the new quantum gauge are 
obtained from the pair of superposed geometries in the old 
quantum gauge by performing different diffeomorphisms on the 
two geometries. Since each diffeomorphism has no effect on the 
geometry, as discussed in section 5, we may expect that this 
transformation on the superposed geometries also would not 
affect any physical phenomenat. Indeed, careful 
analysis [23,20] shows that in the new quantum gauge also the 
same effect occurs, although the mathematical analysis now is 
very different.  

In general, if there is a quantum superposition of gravitational 
fields, by a quantum diffeomorphism, or simply a q-
diffeomorphism, I mean performing different diffeomorphisms on 
the
superposed gravitational fields. Then the above two quantum 
gauges are related by a 
q-diffeomorphism performed on the quantized gravitational field. 
Because in bringing both metric coefficients to the same form in 
$U$, it is necessary to perform two different diffeomorphisms on 
them. These two diffeomorphisms transform the space-time 
points in $U$ differently. Then, as already mentioned, the above 
mentioned effect is invariant under this q-diffeomorphism. 

I postulate now that all physical effects are invariant under all 
q-diffeomorphisms. This suggests a generalization of the usual 
principle of general covariance for the classical gravitational field 
to the following {\it principle of quantum general covariance} in 
quantum gravity: The laws of physics should be covariant under 
q-diffeomorphisms.  

On the other hand, the usual principle of general covariance 
requires 
covariance of the laws of physics under classical diffeomorphisms, 
or c-diffeomorphisms. A c-diffeomorphism is a diffeomorphism 
that is the same for all the superposed gravitational fields, and is 
thus a special case of a q-diffeomorphism. Therefore, the above 
principle of quantum general covariance generalizes the usual 
general covariance due to Einstein. Under a c-diffeomorphism, a 
given space-time point is mapped to the same space-time point 
for 
all of the geometries corresponding to the superposed 
gravitational fields. This is consistent 
with regarding the space-time manifold as real, i.e. a four 
dimensional ether. So, if we restrict to 
just c-diffeomorphism freedom, 
space-time 
may be regarded as objective and real, as already shown in 
section 5.

But the space-time points
associated with each of the superposed gravitational fields,
which are defined above in a
c-diffeomorphism invariant manner,
transform
differently under a q- 
diffeomorphism. This means that in
quantum gravity space-time 
points have no invariant meaning. However, protective 
observation suggests that quantum states are real [3]. 
Consequently, the space-time manifold, which appears to be 
redundant, may 
be discarded, and we may deal directly with the
quantum states of the gravitational field. 

It is the quantum uncertainty in the gravitational field which 
makes points of space-time meaningless. Should we quantize the 
set so that cardinality is itself is uncertain? Since I showed [1] that 
cardinality is a physical and geometrical property, it would seem 
reasonable to quantize it.
But then the curve 
$\gamma$ in 
the gravitational phase operator
(4.1) cannot be meaningfully defined as a curve in space-time. 
The 
resolution of this difficulty may be expected to lead us to a 
quantum theory of 
gravity that may be both operational and geometrical.

\vskip .5cm
\noindent{8. CONCLUDING REMARKS}
\vskip .5cm

The above analysis suggests a philosophical 
principle which may be schematically be expressed as
$${\rm Ontology = Geometry = Physics.} \eqno(8.1)$$

\noindent
The last equality has not been achieved yet by
physicists because we do not have quantum gravity. But it is 
proposed here as a philosophical principle which 
should ultimately
be satisfied by a physical theory.

In relation to the hole argument, described in section 5, this 
principle implies that the
points of space-time become real in classical physics in virtue 
of the
geometrical relations between them. The operational 
determination 
of an
event as an intersection of material world-lines (e.g. lightening
striking the railway track a la Einstein) is a way of observing it. 
This is like, for example, observing 
temperature
by a thermometer. There are many different thermometers which 
may be used, but the concept of temperature is independent of 
them and may be defined as the average kinetic energy of the 
molecules at a given location. Relating a concept to experiments 
does not deny the possibility of an intrinsic meaning to that 
concept.

But in quantum physics, space-time points are not meaningful [1]. 
This is particularly so when gravity is quantized, as argued in 
section 7, above. The symmetry ontology which was proposed in 
section 4, however, suggests (8.1) by providing a link between 
ontology, geometry and physics, as I shall argue now. The 
symmetries of 
the laws of physics are universal in the sense that they are the 
same 
for all laws and for all physical systems governed by them. They 
are 
independent of the particular spaces on which they 
act depending on the particular systems. The symmetries should 
therefore be used to construct the geometry, which should be 
universal. The conserved quantities implied by these symmetries, 
via 
Noether's theorem, is the ``stuff'' of the universe, and may be 
called 
real. The interactions depend on these conserved quantitites and 
elements of the symmetry group as we saw in section 4.  {\it 
Symmetry is 
destiny}.

The question then arises as to how space-time may be obtained 
from 
the symmetry group. I believe that this is due to a common 
property 
of all interactions which picks out a subgroup of the symmetry 
group, which may be called the isotropy group. Then space-time 
emerges as 
the coset space of the isotropy subgroup. This is true not only of 
classical space-time but also the quantum space-time introduced 
in 
section 6. In relation to space-time, this common property may 
therefore be called the locality 
of 
the interactions. That all interactions should possess this common 
property suggests that they should all be unified into a single 
interaction.

In pursuing this new unified theory of all interactions, it may be 
worthwhile to 
keep in mind the following statement due to Einstein, paraphrased 
by Bergmann [24], which I learned from Abner Shimony: ``... a 
systematization of the experimental facts is by itself not yet a 
physical theory and ... in many respects the theoretical physicist is 
a 
philosopher
in workingman's clothes".

\vskip .5cm
\noindent{ACKNOWLEDGEMENTS}
\vskip .5cm

I thank H. R. Brown for useful discussions.
\vskip .5cm
\noindent{REFERENCES}
\vskip .5cm
\item{[1]}J. Anandan, Foundations of Physics, {\bf 10,} 601 
(1980).
\item{[2]}J. Anandan, Foundations of Physics, {\bf 21,} 1265 
(1991).
\item{[3]} Y. Aharonov, J. Anandan, and L. Vaidman, Phys. Rev. A 
{\bf 47,} 4616 (1993); Y. Aharonov and L. Vaidman, Phys. Lett. A 
{\bf 178,} 38 
(1993);  J. Anandan, Foundations of Physics 
Letters {\bf 6,} 503 (1993).
\item{[4]}J. Ehlers, F. A. E. Pirani, and A. Schild, in {\it Papers in 
Honour of J. L. Synge}, edited by L. O'Raifeartaigh (Clarendon 
Press, 
Oxford 1972).
\item{[5]} J. Anandan in {\it Quantum Theory and Gravitation}, 
edited by A. R. Marlow (Academic Press, New York 1980), p. 157.
\item{[6]} J. Anandan, Phys. Rev. D {\bf 15,} 1448 (1977).
\item{[7]}J. Anandan, Nuov. Cim. A {\bf 53,} 221
(1979).
\item{[8]} D. Greenberger, Ann. Phys. {\bf 47,} 116 (1986); D. 
Greenberger and A. W. Overhauser, Sci. Am. {\bf 242,} 66 (1980).
\item{[9]}S. Kobayashi and K. Nomizu, {\it Foundations of 
Differential 
Geometry} (John Wiley, New York 1963). 
\item{[10]} C. N. Yang in {\it Schrodinger: Centenary Celebration of 
a Polymath}, edited by C. W. Kilmster (CUP, 1987).
\item{[11]}E. Schrodinger, Z. f. Phys. 12, 13 (1922).
\item{[12]}C. N. Yang, Phys. Rev. Lett. {\bf 33,} 445 (1974).
\item{[13]}J. Anandan, Phys. Rev. D {\bf 53,} 779 (1996).
\item{[14]} Y. Aharonov and D. Bohm, Phys. Rev. {\bf 115,} 485 
(1959).
\item{[15]} J. Anandan, Phys. Lett. A {\bf 195,} 284 (1994).
\item{[16]} See, for example, reference 7, p. 244.
\item{[17]} J. Stachel in {\it Proceedings of the Fourth Marcel 
Grossmann Meeting on General Relativity}, R. Ruffini, ed. (Elsevier, 
Amsterdam, !986), p.1857-1862; J. Stachel in {\it General 
Relativity and Gavitation - Proceedings of the 11th International 
Conference on General Relativity and Gravitation}, M. A. H. 
MacCallum, ed. (Cambridge Univ. Press, 1987), p. 200-208; J. 
Stachel in {\it Einstein Studies, vol. 1: Einstein and the History of 
General Relativity,} D. Howard and J. Stachel, eds. (Birkhauser, 
Boston-Basel-Berlin, 1989), p. 63-100. See also R. Torretti, 
{\it Relativity and 
Geometry} (Pergamon Press, Oxford 1983), 5.6.
\item{[18]} A. Einstein and M. Grossmann, Zeitschr. Math. und 
Phys. {\bf 62,} 225 (1913).
\item{[19]} H. R. Brown, International Studies in the Philosophy of 
Science, vol 9, 235-253, (1995).
\item{[20]} Y. Aharonov and L. Susskind, Phys. Rev. {\bf 158,} 
1237 (1967).
\item{[21]} Y. Aharonov and J. Anandan, in preparation.
\item{[22]} J. Anandan, J. of Gen. Rel. and Grav. {\bf 26,} 125 
(1994).
\item{[23]} Y. Aharonov and J. Anandan, Phys. Lett. A {\bf160,} 
493 (1991); J. Anandan, Phys. Lett. A {\bf164,} 369 (1992).
\item{[24]} P.G. Bergmann, {\it Basic Theories of Physics, vol. 1}, 
preface
(Prentice-Hall, New York, 1949).
\bye